\documentstyle[preprint,aps]{revtex}
\input epsf.tex
\def\DESepsf(#1 width #2){\epsfxsize=#2 \epsfbox{#1}}
\begin{document}
\preprint{\vbox{\hbox{}}}
\draft
\title{Constraints on the phase $\gamma$ and new physics from $B\to K\pi$ Decays
}
\author{
Xiao-Gang He, Chien-Lan Hsueh and Jian-Qing Shi}

\address{
Department of Physics, National Taiwan University, Taipei, Taiwan, 10617
}

\date{May 1999}
\maketitle
\begin{abstract}
Recent results from CLEO on $B\to K\pi$ indicate that 
the phase $\gamma$ may be substantially different from 
that obtained from other
fit to the KM matrix elements in the Standard Model.
We show that $\gamma$ extracted using
$B\to K\pi, \pi\pi$ is sensitive to
new physics occurring at loop level. It provides a powerful method to
probe new physics in electroweak penguin interactions. 
Using effects due to anomalous gauge couplings as an example, 
we show that within the allowed ranges for these 
couplings information about
$\gamma$ obtained from $B\to K \pi, \pi\pi$ can be very different from
the Standard Model prediction.
\end{abstract}

The CLEO collaboration has recently measured the four
 $B\to K\pi$ branching ratios
with\cite{1},
$Br(B^\pm \to \pi^\pm K^0) = (1.82^{+0.46}_{-0.4}\pm 0.16)
\times 10^{-5}$, $Br(B^\pm \to K^\pm \pi^0) =
(1.21^{+0.30+0.21}_{-0.28-0.14})\times 10^{-5}$, $Br(B\to K^\pm \pi^\mp) = 
(1.88^{+0.28}_{-0.26}\pm 0.13)\times 10^{-5}$ and
$Br(B\to K^0 \pi^0) = (1.48^{+0.59+0.24}_{-0.51-0.33})\times 10^{-5}$.
It is suppressing that these branching ratios turn out to be close to each 
other because
naive expectation
of strong penguin dominance would give $R=Br(
K^\pm \pi^0)/Br(K \pi^\pm)\sim 1/2$ and model calculations for $Br(B^0\to K^0 \pi^0)$
would obtain a much smaller value.
The closeness of the branching ratios with charged mesons in the final states 
may be
an indication of large interference effects of tree, strong and
electroweak penguin interactions\cite{2}. 
It has been shown that 
using information from these decays and $B^\pm \to \pi^\pm \pi^0$ decays, the
phase angle $\gamma$ of the KM matrix can be constrained\cite{3} and
determined\cite{4a} in the Standard Model (SM).
Using the present central values for 
these branching ratios  we find that the constraint obtained on $\gamma$
may have potential problem with
$\gamma= (59.5^{+8.5}_{-7.5})^\circ$ obtained from other constraints$\cite{5}$.

If there is new physics beyond the SM
the situation becomes complicated\cite{6,7}.
It is not possible to 
isolate different new physics sources in the most general case.
However, one can extract important
information for the class of models where significant 
new physics effects only show up 
at loop levels for B decays\cite{6,7}.
In this paper we study 
how new physics of the type described
above can affect the results using anomalous three gauge boson couplings as
an example for illustration.

New physics due to anomalous three gauge boson couplings is a perfect
example of models where new physics effects only appear at loop level in B 
decays.
Effects due to anomalous couplings do not appear
at tree level for B decays to the lowest order, 
and they do not affect CP violation and mixings
in $K^0-\bar K^0$ and $B-\bar B$ systems 
at one loop level. Therefore they do not
affect the fitting to the KM parameters obtained in Ref. \cite{5}.
However they affect
the constraint on and determination of
$\gamma$ using experimental results from $B \to K \pi, \pi\pi$,
and affect the $B$ decay branching ratios.

The effective Hamiltonian $H_{eff} = (G_F/\sqrt{2})[V_{ub}^*V_{uq}(c_1O_1+c_2O_2)
-V_{tb}^*V_{tq}\sum_{i=3-10} c_iO_i]$ responsible for B decays have been studied by many
authors\cite{8}. 
We will use the values of $c_i$ obtained for the SM in Ref.\cite{9} with

\begin{eqnarray}
&&c_1=-0.313,\;\;c_2=1.150,\;\;c_3=0.017,\;\;c_4=-0.037,\;\;c_5=0.010,\;\;c_6=-0.046,
\nonumber\\
&&c_7=-0.001\alpha_{em},\;\;c_8=0.049\alpha_{em},\;\;c_9=-1.321\alpha_{em},\;\;
c_{10}=0.267\alpha_{em}.
\end{eqnarray}
The Wilson Coefficients are modified when anomalous couplings are included. 
They will generate non-zero $c_{3-10}$\cite{10}.
Their effects on
$B\to K \pi$ mainly come from $c_{7-10}$. 
To the leading order in QCD corrections, the new contributions to 
$c_{7-10}^{AC}$ 
due to various anomalous couplings are given by,

\begin{eqnarray}
c^{AC}_7/\alpha_{em}&=&-0.287\Delta \kappa^\gamma - 0.045\lambda^\gamma +1.397\Delta g_1^Z
-0.145g_5^Z,\nonumber\\
c^{AC}_8/\alpha_{em}&=&-0.082\Delta \kappa^\gamma - 0.013\lambda^\gamma +0.391\Delta g_1^Z - 0.041g_5^Z
,\nonumber\\
c^{AC}_9/\alpha_{em}&=&-0.337\Delta \kappa^\gamma -0.053\lambda^\gamma -5.651
\Delta g_1^Z + 0.586g_5^Z ,\nonumber\\
c^{AC}_{10}/\alpha_{em}&=&0.069\Delta \kappa^\gamma +0.011\lambda^\gamma +1.143
\Delta g_1^Z - 0.119g_5^Z.
\end{eqnarray}
In the above we have used a cut off $\Lambda = 1$ TeV for terms proportional to $\Delta \kappa^\gamma$
and $\Delta g_1^Z$. 
Contributions from other anomalous couplings
are suppressed by additional factors of order 
$(m_b^2, m_B^2)/m_W^2$ which can be safely neglected.
There are constraints on the anomalous gauge couplings\cite{11,13}. 
LEP experiments obtain\cite{13},
$-0.217<\Delta \kappa^\gamma < 0.223$, $-0.158 < \lambda^\gamma < 0.074$, and
$-0.113 <\Delta g_1^Z < 0.126$ at the 95\% confidence level. Assuming 
that $g_5^Z$ 
is the same order as $\Delta g_1^Z$, it is clear that the largest
possible 
contribution may come from $\Delta g^Z_1$. In our later discussions we  
consider $\Delta g^Z_1$ effects only.

To see 
possible deviations from the SM predictions for $B\to K\pi$ data, we
carried out a calculation using factorization following Ref. \cite{ali}
with $V_{us} = 0.2196$, $V_{cb} \approx -V_{ts} = 0.0395$, 
$|V_{ub}| = 0.08 V_{cb}$\cite{18}.
The branching ratios as functions of $\gamma$
are shown in Fig. 1. In this figure we used
$m_s = 100$ MeV which is at the middle of the range from lattice 
calculations\cite{lattice} and the number of colors $N_c = 3$.
Since penguin dominates the branching ratio for $Br(B^+\to K^0\pi^+)$ which 
is insensitive to $\gamma$, we normalize
the branching ratios to $Br(B^+\to K^0 \pi^+)$ to reduce possible
uncertainties in the overall normalization of form factors involved.

From Fig. 1, we see that the central values for the branching ratios for the
ones with at least one charged meson in the final states require the
angle $\gamma$ to be within $75^\circ$ to $80^\circ$ rather 
than the best fit value $\gamma_{best} = 59.5^\circ$ in Ref.\cite{5}. 
Larger $\gamma$ is also indicated by other rare
B decay data\cite{13a}. 
When effects due to $\Delta g^Z_1$ is included the situation can be relaxed.  
The effects of $\Delta g^Z_1$ on $B^+\to K^0\pi^+, K^+\pi^-$ 
are very small, but are significant for $B^+\to K^+\pi^0$ and $B\to K^0\pi^0$.
With positive $\Delta g^Z_1$ in its allowed range, 
it is possible for the relative ratios of $Br(B^+\to K^+ \pi^0)$ to the other
charged modes to be in agreement with data for $\gamma=\gamma_{best}$.
We note that $\Delta g^Z_1$ does not affect the ratio
$Br(B^0\to K^+\pi^-)/Br(B^+\to K^0\pi^+)$. Its experimental value
prefers $\gamma$ to be close to $75^\circ$.
Of course we also note that this situation can be improved by treating
$N_c$ as a free parameter to take into account certain non-factorizable effects.
We find that with $N_c\approx 1.35$, the central experimental values for
the branching ratios of B decays into charged mesons in the final states
can be reproduced for $\gamma = \gamma_{best}$.
It is not possible to bring $Br(B^0\to K^0\pi^0)$ up to the experimental 
central value even with allowed $\Delta g^Z_1$ and reasonable values for
$N_c$ and $m_s$.

If the present experimental central values will persist and factorization
approximation with $N_c =3$ is valid, new physics may be 
needed. Needless to say that we have to wait for more accurate data to draw
firmer conclusions. 
Also due to our inability to reliably 
calculate the hadronic matrix elements, one should be careful in drawing 
conclusions with factorization calculations. 
However, we would like to point out that data on rare B to $K\pi$ decays
may provide a window to look for new physics beyond the SM.
Of course, to have a better understanding of the situation one needs to find 
methods which are able 
to extract $\gamma$ in a model independent way and in the
presence of new physics. In the following we will analyze some rare
B to $K\pi$ decay data in a more model independent way.

Model independent constraint on $\gamma$ can  be obtained using branching ratios for
$B^\pm\to K\pi$ and $B^\pm\to \pi\pi$ from symmetry considerations.
This method would only need information from charged B decays to $K\pi$ and
$\pi\pi$, and therefore this method is not affected by the
uncertainties associated with neutral B decays to $K\pi$ modes. 
Using SU(3) relation and factorization estimate  for
SU(3) breaking effect among $B^\pm \to K\pi$ and $B^\pm \to \pi\pi$, one 
obtains\cite{3,4a}

\begin{eqnarray}
&&A(\pi^+ K^0) + \sqrt{2}A(\pi^0 K^+)
=\epsilon A(\pi^+ K^0)e^{i\Delta \phi}
{e^{i\gamma}-\delta_{EW} \over
1-\delta_{EW}'},\nonumber\\
&&
\delta_{EW} = -{3\over 2} {|V_{cb}||V_{cs}|\over |V_{ub}||V_{us}|}{c_{9}+c_{10}\over c_1+c_2},
\;\;\;\;\delta_{EW}' = {3\over 2} {|V_{tb}||V_{td}|\over |V_{ub}||V_{ud}|}e^{i\alpha}
{c_{9}+c_{10}\over c_1+c_2},\nonumber\\
&&\epsilon = \sqrt{2} {|V_{us}|\over |V_{ud}|} {f_K\over f_\pi}
{|A(\pi^+\pi^0)|\over |A(\pi^+ K^0)|},
\label{su3}
\end{eqnarray}
where $\Delta \phi$ is the difference of the final state elastic re-scattering phases
$\phi_{3/2, 1/2}$ for $I=3/2,1/2$ amplitudes. 
For $f_K/f_\pi = 1.22$ and 
$Br(B^\pm \to \pi^\pm \pi^0) = (0.54^{+0.21}_{-0.20}\pm 0.15)\times 10^{-5}$, 
we obtain $\epsilon = 0.21 \pm 0.06$.

The parameter $\delta_{EW}'$ is of order $c_{9,10}/c_{1,2}$ which 
is much smaller than one and will be 
neglected in our later discussions.
$\delta_{EW}$ is a true measure of electroweak 
penguin interactions in hadronic 
B decays and provides an easier probe of these interactions compared with
other methods\cite{ew}.
In the above contributions from $c_{7,8}$ have been neglected which is safe in the SM
because they are small. 
With anomalous couplings this is 
still a good approximation. In general the contributions
from $c_{7,8}$ may be substantial. In that case the expression becomes more complicated. But
one can always absorb the contribution into an effective $\delta_{EW}$. 
In the SM for $r_v=|V_{ub}|/|V_{cb}| = 0.08$ and $|V_{us}| = 0.2196$\cite{18}, 
$\delta_{EW} = 0.81$. Smaller $r_v$ implies larger $\delta_{EW}$. Had we used 
$r_v=0.1$, $\delta_{EW}$ would be 0.65 as in Ref. \cite{3,4a}. 
With anomalous couplings, we find

\begin{eqnarray}
\delta_{EW} = 0.81(1+0.26\Delta \kappa^\gamma +0.04\lambda^\gamma+
4.33\Delta g_1^Z -0.45g_5^Z).
\end{eqnarray}
The value for $\delta_{EW}$ can be different from the SM prediction. It is most
sensitive to $\Delta g_1^Z$.
Within the allowed range of $-0.113 < \Delta g_1^Z < 0.126$ \cite{13},
$\delta_{EW}$ 
can vary in the range
$0.40\sim 1.25$.

Neglecting small tree contribution to $B^+\to \pi^+K^0$, one obtains
\begin{eqnarray}
&&\cos\gamma =\delta_{EW} -
{(r^2_++r^2_-)/2 -1 - \epsilon^2(1-\delta_{EW}^2) 
\over 2 \epsilon (\cos \Delta \phi
+ \epsilon \delta_{EW})},
\label{gamma1}
\end{eqnarray}
\begin{eqnarray}
&&r^2_+-r^2_- = 4\epsilon \sin \Delta \phi \sin \gamma,
\label{gamma2}
\end{eqnarray}
where
$r_\pm^2 = 4Br(\pi^0 K^\pm)/[Br(\pi^+ K^0)
+ Br(\pi^- \bar K^0)] = 1.33\pm 0.45$. 

If SU(3) breaking effect is indeed represented by the last equation in
(\ref{su3}), and tree contribution to $B^\pm \to \pi^\pm K$ is small, 
information about $\gamma$ and 
$\delta_{EW}$ obtained are free from uncertainties associated with hadronic 
matrix elements.
Possible SU(3) breaking effects have been estimated and shown to be 
small\cite{3,4a,15}.
The smallness of the tree contribution to  $B^\pm \to K\pi^\pm$ is 
true in factorization approximation and can be checked 
experimentally\cite{16}. 
The above equations can be tested in the future. 
We will assume the validity of Eq. (\ref{gamma1}) and
study how information on
$\gamma$ obtained from $B\to K\pi,\pi\pi$ decays depends on $\delta_{EW}$.

The relation between $\gamma$ and $\delta_{EW}$ is complicated. However
it is interesting to note that 
even in the most general case, bound on $\cos\gamma$ can be 
obtained. For $\Delta = (r^2_+ + r^2_-)/2 -1-\epsilon^2(1-\delta_{EW}^2)>0$, we have

\begin{eqnarray}
&&\cos\gamma \le \delta_{EW}- {\Delta\over 2\epsilon (1+\epsilon \delta_{EW})}, \;\;
\mbox{or}\;\;\;
\cos\gamma \ge \delta_{EW}-{\Delta \over 2\epsilon (-1+\epsilon \delta_{EW})}.
\label{bound}
\end{eqnarray}
The sign of $\Delta$ depends on $r^2_\pm$, $\epsilon$ and $\delta_{EW}$. As long
as $r^2_\pm >1.07$, $\Delta$ is larger than zero at the
 90\% C.L. in the allowed range for $\epsilon$ and  
any value for $\delta_{EW}$. For smaller $r^2_\pm$, $\Delta$ can change
sign depending on $\delta_{EW}$.  
For $\Delta <0$, the bounds are given by replacing $\le$, $\ge$ by
$\ge$, $\le$ in the above equations, respectively. 
We remark that if $r^2_\pm <1$, one can also use the method in 
Ref. \cite{19} to constrain $\gamma$.
The above bounds become exact solutions for $\cos \Delta \phi =1$ 
and $\cos\Delta \phi = -1$, respectively. For $\epsilon <<1$, one obtains the 
bound $|\cos\gamma - \delta_{EW}| \ge (r^2_+ + r^2_- -2)/(4\epsilon)$ in
Ref. \cite{3}.

We will
use the central value for $\epsilon$ and vary $r^2_\pm$ in our numerical 
analysis to illustrate how information on $\gamma$ and its dependence on
new physics through $\delta_{EW}$ can be obtained.
The bounds on $\cos\gamma$ are shown in Fig. 2 
by the solid curves
for three representative cases: 
a) Central values for $\epsilon$ and $r^2_\pm$;
b) Central values for $\epsilon$ and $1 \sigma$ upper bound $r^2_\pm=1.78$;
 and c) Central value for $\epsilon$ and $1 \sigma$ lower bound $r^2_\pm = 0.88$.
For cases a) and b) $\Delta >0$, and for case c) $\Delta <0$.

The bounds with $|\cos\gamma| \le 1$ for a), b) and c) are indicated by   
the curves (a1, a2), (b) 
and (c1, c2), respectively.
For cases  a) and c) there are two allowed 
regions, 
the regions below (a1, c1) and the regions above (a2, c2).
For case b) the allowed range is below (b).
When $r^2_\pm$ decreases from $1 \sigma$ upper bound to $1 \sigma$ lower bound,
one of the boundaries goes up from (b) to (a1) then moves to (c2). 
And the other boundary for case b) which is
outside  the range moves to (a2) then goes down to (c1).
In case a), for $\delta_{EW} = 0.81 (0.65) $ we find $\cos\gamma <0.18 (0.015)$ 
which is
way below the value corresponding to $\cos\gamma_{best} \approx 0.5$. 
With $\Delta g^Z_1 = 0.126$, $\cos\gamma$ can be close to 0.5. 
For larger $r^2_\pm$ the situation is worse. This can be seen from 
the curves for case b) where 
$\cos\gamma <0$ for $\delta_{EW}$ up to 1.5. 
For smaller $r_\pm^2$ the situation is better as can be seen from 
case c).
In this case there are larger allowed ranges.
$\gamma \approx \gamma_{best}$ can be accommodated even by the SM.

When the decay amplitudes for $B^\pm\to K\pi$,
$B^\pm \to \pi^\pm \pi^0$ and the rate asymmetries for these decays
are determined to a good accuracy,
$\gamma$ can be determined using Eq. (\ref{su3}) and its conjugated form.
The original method\cite{18a} using similar equations without the correction
$\delta_{EW}$ from
electroweak penguin is problematic because the correction is large
\cite{20}. Many variations involving other
decay modes have been proposed to remove electroweak penguin 
effects\cite{21}.
Recently it was realized\cite{4a} that the difficulties associated with
 electroweak penguin
interaction can be calculated in terms of the quantity
$\delta_{EW}$.

This method again crucially depends on the value of
$\delta_{EW}$. The solution of $\cos\gamma$ corresponds to the
solution of a fourth order polynomial in $\cos\gamma$.
Using Eqs. (\ref{gamma1}) and (\ref{gamma2}), we have

\begin{eqnarray}
(1-\cos^2\gamma) [ 1-({\Delta \over 2\epsilon(\delta_{EW} -\cos \gamma)}
-\epsilon \delta_{EW})^2] - {(r^2_+-r^2_-)^2\over 16 \epsilon^2} = 0.
\end{eqnarray}
The solutions depend on the values of $r^2_\pm$ and $\epsilon$ which are 
not determined with sufficient accuracy at present. 
To have some idea about the details, we analyze 
the solutions of $\cos \gamma$ as a function of $\delta_{EW}$
for the three cases discussed earlier 
with a given value for the asymmetry $A_{asy} = (r^2_+-r^2_-)/(r^2_++r^2_-)$.
In Fig. 2 we show the solutions with 
 $A_{asy} = 15\%$ for
illustration. The actual value to be used for practical analysis has to be determined
by experiments. 
The solutions for the three cases a), b) and c) are indicated by the 
dashed, dot-dashed and dotted curves in Fig. 2.
In general there are four solutions, but not all of them are physical ones satisfying
$|\cos \gamma| <1$. 

For case a), 
two solutions are allowed with $\delta_{EW}$.
To have $\cos\gamma >0$ $\delta_{EW}$ has to be larger than
0.7. Whereas $\cos\gamma \approx \cos\gamma_{best}$ would require 
$\delta_{EW}$ to be larger
than 1.2 which can not be reached in the SM, but is possible for non-zero
 $\Delta g^Z_1$ in its allowed range.
With smaller $r^2_\pm$, 
 $\cos\gamma >0$ can be solution with smaller $\delta_{EW}$ and 
can even have $\cos\gamma = \cos\gamma_{best}$. This can be seen 
from the dotted
curves in Fig. 2 for case c).
For larger $r^2_\pm$ in order to have solutions, larger $\delta_{EW}$ is
required. For case b) $\delta_{EW}$ must be larger than 1.4 in order to 
have solutions.
These regions can not be reached by SM, nor by the model with 
$\Delta g^Z_1$ in the allowed range. 

We also analyzed how the solutions change with the asymmetry $A_{asy}$. 
With small $A_{asy}$, the solutions are close to the bounds. When $A_{asy}$
increases, the solutions move away from the bounds. 
The solutions below the bounds (a1), (b) and (c1) shift towards the right, 
and the bounds (a2) and (c2) move towards the left. 
In all cases the solutions with $|\cos \gamma| \approx 1$ 
become more sensitive
to $\delta_{EW}$ and $|\cos\gamma|$ becomes smaller as $A_{asy}$ increases. 
In each case discussed the solutions, except the ones close to 
$|\cos\gamma| =1$, in models with non-zero $\Delta g_1^Z$ 
can be very different from those in the SM.
It is clear that 
important information about $\gamma$ and about new physics contribution 
to $\delta_{EW}$ can be 
obtained from $B^\pm\to K\pi, \pi\pi$ decays.

We conclude that 
the branching ratios of $B\to K\pi$ decays are sensitive to new physics at loop level. 
The bound on $\gamma$, extracted using 
the central branching ratios for $B^\pm \to K\pi$
and information from $B^\pm \to \pi^\pm \pi$,  
is different from that obtained from other experimental data.
New physics, such as anomalous gauge couplings, can improve the situation.
Similar analysis can be applied to any other model where new physics 
contribute to electroweak penguin interactions. 
The decay modes, $B^\pm\to K\pi, \pi\pi$ will be measured 
at various B factories with improved error bars. 
The Standard Model and models beyond will be tested in the near future.

\noindent {\bf Acknowledgments:}

This work was partially supported by National Science Council of
R.O.C. under grant number  NSC 88-2112-M-002-041.

\begin{figure}[htb]
\centerline{ \DESepsf(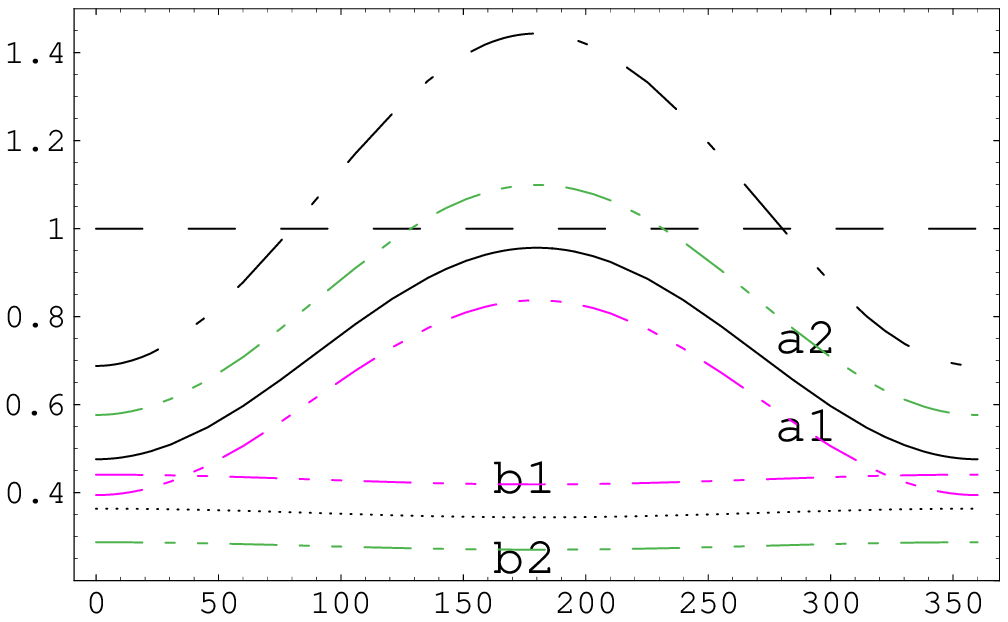 width 10cm)}
\smallskip
\caption {
CP-averaged branching ratios normalized to $Br(B^+\to K^0\pi^+)$ vs. $\gamma$
 for $B^+\to K^0 \pi^+$ (dashed), 
$B^+\to K^+\pi^0$ (solid), $B^0\to K^+ \pi^-$ (dot-dashed), and
$B^0\to K^0 \pi^0$ (dotted) for the 
SM with $m_s = 100$ MeV. The curves $a_1,\;b_1$ and $a_2,\;b_2$ are 
for $\Delta g_1^Z$ equal to $-0.113$ and $0.126$, respectively. 
}
\label{kpi-br}
\end{figure}

\begin{figure}[htb]
\centerline{ \DESepsf(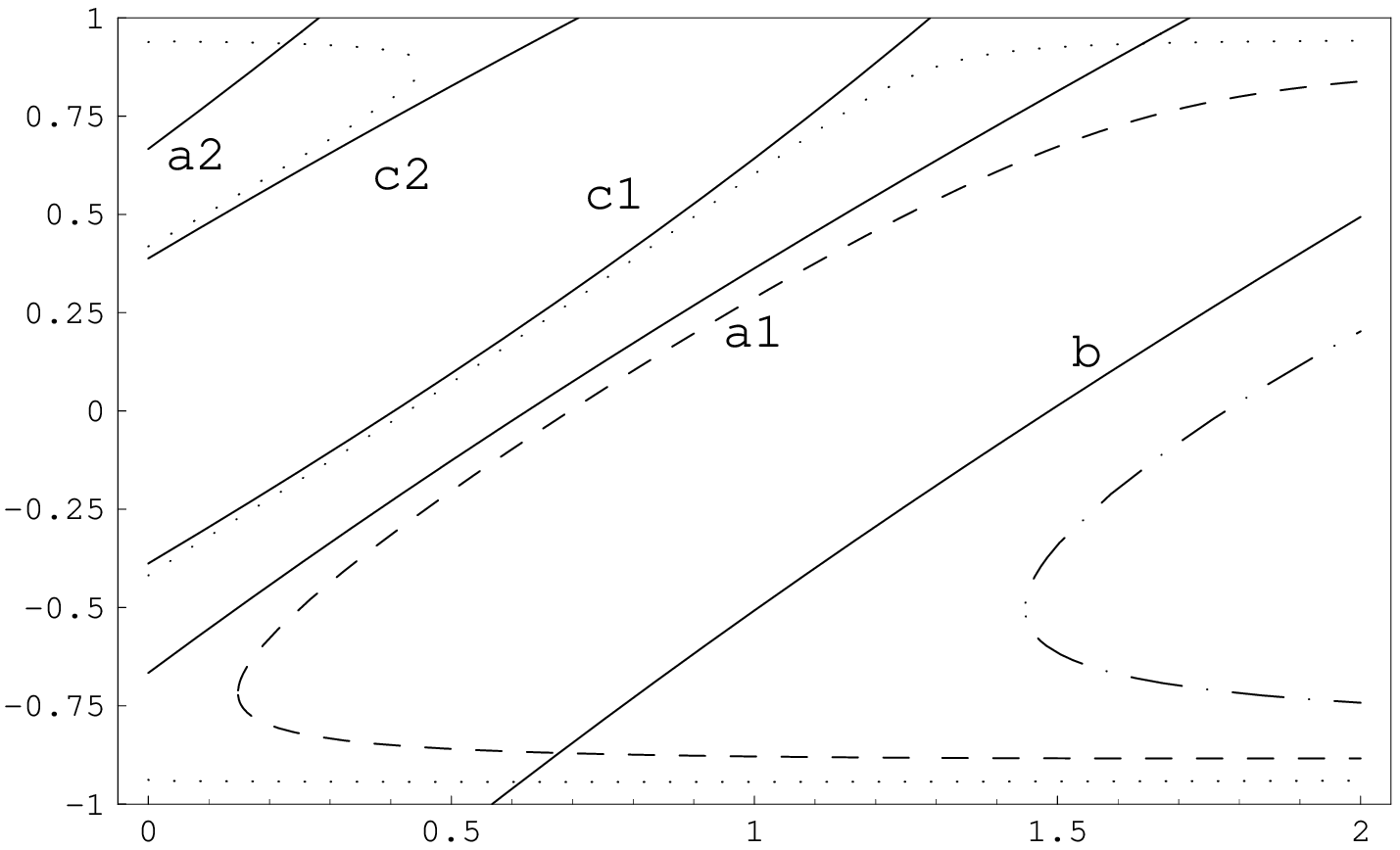 width 10cm)}
\smallskip
\caption {
Bounds on and (solutions for) $\cos\gamma$ vs. $\delta_{EW}$.
The curves a1, a2 (dashed), b (dot-dashed),  and c1,c2 (dotted)
 are the 
bounds (solutions) on (for)
 $\cos\gamma$ as functions of $\delta_{EW}$ for the
three cases a), b) and c) described in the text, respectively. 
}
\label{kpi-angle}
\end{figure}

\vspace{1cm}


\begin{thebibliography}{99}

\bibitem{1} CLEO Collaboration, Y. Kwon et al., CLEO CONF 99-14.

\bibitem{2} N. Deshpande et al, Phys. Rev. Lett. {\bf 82},2240(1999).


\bibitem{3} M. Neubert and J. Rosner, Phys. Lett. {\bf B441}, 403(1998).

\bibitem{4a} M. Neubert and J. Rosner, Phys. Rev. Lett. {\bf 81}, 5074(1998);
M. Neubert, JHEP 9902:014 (1999).

\bibitem{5} F. Parodi, P. Roudeau and A. Stocchi, e-print hep-ex/9903063.

\bibitem{6} X.-G. He, Phys. Rev. {\bf D53}, 6326(1996);
X.-G. He, CP Violation, e-print hep-ph/9710551, in proceedings
of 
"CP violation and various frontiers in Tau and other systems", CCAST, Beijing, 1997.

\bibitem{7} N. Deshpande, B. Dutta and S. Oh, Phys. Rev. Lett. {\bf 77}, 4499(1996);
A. Cohen et al., Phys. Rev. Lett. {\bf 78}, 2300(1997);
Y. Grossman, Y. Nir and M. Worah, Phys. Lett. {\bf B407}, 307(1997);
M. Ciuchini et. al., Phys. Rev. Lett. {\bf 79}, 978(1997).

\bibitem{8} M. Lusignoli, Nucl. Phys. {\bf B325}, 33(1989);
A. Buras, M. Jamin and M. Lautenbacher, Nucl. Phys. {\bf B408}, 209(1993);
M. Cuichini et. al., Nucl. Phys. {\bf B415}, 403(1994).

\bibitem{9} N. Deshpande and X.-G. He, Phys. Lett. {\bf B336}, 471(1994).

\bibitem{10} X.-G. He and B. McKellar, Phys. Rev. {\bf D51}, 6484(1995); 
X-G. He, Phys. Lett. {\bf B460}, 405(1999).

\bibitem{11} S.-P. Chia, Phys. Lett. {\bf B240}, 465(1990);
K. A. Peterson, Phys. Lett. {\bf B282}, 207(1992);
K. Namuta, Z. Phys. {\bf C52}, 691(1991);
T. Rizzo, Phys. Lett. {\bf B315}, 471(1993);
X.-G. He, Phys. Lett. {\bf B319}, 327(1994);
X.-G. He and B. McKellar, Phys. Lett. {\bf 320}, 168(1994);
S. Dawson and G. Valencia, Phys. Rev. {\bf D49}, 2188(1994);
G. Baillie, Z. Phys. {\bf C61}, 667(1994);
G. Burdman, Phys. Rev. {\bf D59}, 035001(1999).

\bibitem{13} ALEPH Collaboration, ALEPH 99/072, CONF 99/046.

\bibitem{ali} A. Ali, G. Kramer and C.-D. Lu, Phys. Rev. {\bf D58}, 094009(1998).

\bibitem{18} Particle Data Group, Eur. Phys. J. {\bf C3}, 1(1998).

\bibitem{lattice} For a recent review see, Sinead Ryan, e-print hep-ph/9908386.

\bibitem{13a} X.-G. He, W.-S. Hou and K.-C. Yang, Phys. Rev. Lett. {\bf 83},
1100(1999).


\bibitem{ew} R. Fleischer, Phys. Lett. {\bf B321}, 259(1994);
N. Deshpande, X.-G. He and J. Trampetic, Phys. Lett. {\bf B345}, 547(1995).

\bibitem{15} A. Buras and R. Fleischer, e-print hep-ph/9810260.

\bibitem{16} M. Gerard and J. Weyers, e-print hep-ph/9711469; D. Delepine et al.,
Phys. Lett. {\bf B429}, 106(1998);
A. Falk et. al., Phys. Rev. {\bf D57}, 4290(1998); D. Atwood and A. Soni. Phys. Rev. 
{\bf D58}, 036005(1998); 
M. Neubert, Phys. Lett. {\bf B422}, 152(1998);
R. Fleischer, Phys. Lett. {\bf B435}, 221(1998); Eur. Phys. J. {\bf C6}, 
451(1999);
M. Gronau and J. Rosner, Phys. Rev. {\bf D57}, 6843(1998);
{\bf D58}, 113005(1998);
B. Block, M. Gronau and J, Rosner, Phys. Rev. Lett. {\bf 78}, 3999(1997);
W.-S. Dai et al., e-print hep-ph/9811226 (Phys. Rev. D in press);
X.-G. He, Eur. Phys. J. {\bf C9}, 443(1999).

\bibitem{19} R. Fleischer and T. Mannel, Phys. Rev. {\bf D57}, 2752(1998).

\bibitem{18a} D. London, M. Gronau and J. Rosner. Phys. Rev. Lett. {\bf 73},
21(1994).
 
\bibitem{20} N. Deshpande and X.-G. He, Phys. Rev. Lett. {\bf 74}, 26 (E: 4099)(1995).

\bibitem{21} N. Deshapande, and X.-G. He, Phys. Rev. Lett. {\bf 75}, 3064(1995);  N. Deshpande, X.-G. He and S. Oh, Z. Phys. {\bf C74}, 359(1997);
M. Gronau et al., Phys. Rev. {\bf D52}, 6374(1995);
A. Dighe, M. Gronau and J, Rosner, Phys. Rev. {\bf D54}, 3309(1996);
M. Gronau and J. Rosnar Phys. Rev. Lett. {\bf 76}, 1200(1996);
R. Fleischer, Int. J. Mod. Phys. {\bf A12}, 2459(1997).


\end{thebibliography}
\end{document}